\documentclass[aps,prb,twocolumn,showpacs,amsmath,amssymb]{revtex4}

\usepackage{graphicx}
\usepackage{dcolumn}
\usepackage{bm}
\begin{document}

\title{Critical currents and vortex dynamics in percolative
superconductors\\ containing fractal clusters of a normal phase}

\author{Yuriy I. Kuzmin}
\email{yurk@mail.ioffe.ru, iourk@yandex.ru}

\affiliation{Ioffe Physical Technical Institute of the Russian
Academy of Sciences, 26 Polytechnicheskaya Street, Saint
Petersburg 194021 Russia}

\date{\today}

\begin{abstract}
The effect of fractal clusters on magnetic and transport properties of
percolative superconductors is considered. The superconductor contains
percolative superconducting cluster, carrying a transport current, and
clusters of a normal phase, which act as pinning centers. A prototype of
such a structure is a high-temperature superconducting (HTS) wire. The
superconducting core of first generation HTS wires is the conglomeration of
superconducting micro-crystallites containing normal-phase clusters inside.
There are the clusters of columnar defects in superconducting layer, which
is the current-carrying element of second generation HTS wire. It is found
that normal-phase clusters can have essential fractal features that affect
the vortex dynamics. The fractal dimension of the boundary of normal-phase
clusters in YBCO films is estimated and the cluster statistics is analyzed.
Depinning and transport of vortices in fractal superconducting structures
are investigated. Transition of the superconductor into a resistive state
corresponds to the percolation transition from a pinned vortex state to a
resistive state when the vortices are free to move. It is revealed that a
mixed state of the vortex glass type is realized in the superconducting
system involved. The fractal distribution of critical currents is derived
and its peculiarities are studied. It is found that there is the range of
fractal dimension where this distribution has anomalous statistical
properties, specifically, its dispersion becomes infinite. The
current-voltage characteristics of superconductors containing fractal
clusters are obtained. Dependencies of the free vortex density on the
fractal dimension as well as the resistance on the transport current are
studied. It is found that the fractality of the cluster boundary intensifies
pinning and thereby raises the critical current. This feature enables the
current-carrying capability of a superconductor to be enhanced without
changing of its chemical composition.
\end{abstract}


\maketitle

\section{Introduction}

An essential feature of the clusters of defects in superconductor
lies in their capability to trap a magnetic flux.
\cite{tonomura,lidmar,dam} Holding in place the vortices driven by
the Lorentz force, such clusters can act as effective pinning
centers. \cite{higuchi,mezzetti} This feature can be used in
making new composite superconducting materials of enhanced
current-carrying capability. \cite{beasley,krusin,tpl2000} The
cluster structure affects the vortex dynamics in superconductors,
especially when clusters have fractal boundary.
\cite{olson,surdeanu,prester,pla2000,prb} In the present paper the
magnetic and transport properties of composite superconductors
with fractal clusters of a normal phase will be considered as well
as the phenomena limiting the current-carrying capability of such
superconductors will be discussed.

\begin{figure}
\includegraphics{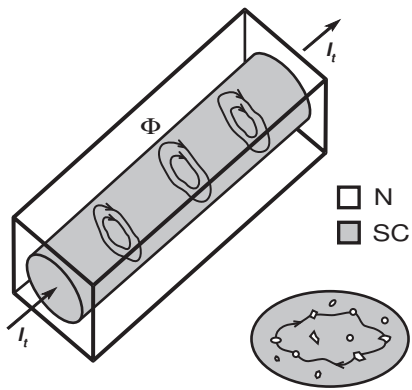}
\caption{\label{fig1} Schematic representation for the first
generation superconducting wire. The superconducting core is
sheathed by the normal metal. The transport current creates the
magnetic flux concentrated along irregular-shaped rings passing
through the normal-phase inclusions. The inset shows the
cross-section of the wire with normal-phase inclusions as well as
one cluster defined as a set of pinning centers, which are united
by the common trapped flux.}
\end{figure}

A prototype of superconductor containing inclusion of a normal
phase is a superconducting wire. The first generation
high-temperature superconducting (HTS) wires are fabricated
following the powder-in-tube technique (PIT). The metal tube is
being filled with HTS powder, then the thermal and deformation
treatment is being carried out. The resulting product is the wire
consisting of one or more superconducting cores sheathed by a
normal metal (Fig.~\ref {fig1}). The sheath endows the wire with
the necessary mechanical (flexibility, folding strength) and
electrical (the possibility to release an excessive power when the
superconductivity will be suddenly lost) properties. At present,
the best results are obtained for the silver-sheathed
bismuth-based composites, which are of practical interest for
energy transport and storage. In view of the PIT peculiarity the
first generation HTS wire has a highly inhomogeneous structure.
\cite{pashitski} Superconducting core represents a dense
conglomeration of BSCCO micro-crystallites containing normal-phase
inclusions inside (see inset in Fig.~\ref{fig1}). These inclusions
primarily consist of a normal metal (silver) as well as the
fragments of different chemical composition, grain boundaries,
micro-cracks, and the domains of the reduced superconducting order
parameter. \cite{fukumoto, suenaga,hlasnik} The volume content of
a normal phase in the core is far below the percolation threshold,
so there is a percolative superconducting cluster that carries the
transport current.

The second generation HTS wires (coated conductors) have
multi-layered film structure consisting of the metal substrate
(nickel-tungsten alloy), the buffer oxide sub-layer, HTS-layer
(YBCO), and the protective cladding made from a noble metal
(silver). Superconducting layer, which carries the transport
current, has the texture preset by the structure of an oxide
sub-layer. In the superconducting layer there are clusters of
columnar defects (see Fig.~\ref{fig2}) that can be created during
the film growth process as well as by the heavy ion bombardment.
\cite{mezzetti,smith} Such defects are similar in topology to the
vortices, therefore they suppress the flux creep so it is possible
to get the critical current up to the depairing value.
\cite{tonomura,indenbom}

\section{Magnetic Flux Trapping in Percolative Superconductors}

\begin{figure}
\includegraphics{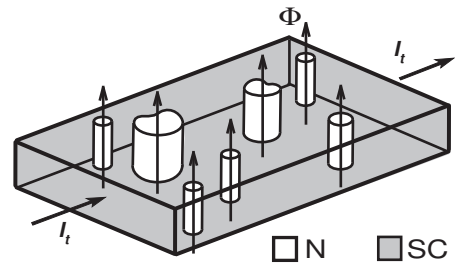}
\caption{\label{fig2} Superconducting film with columnar clusters
of a normal phase. Such a film represents the current-carrying
element of the second generation superconducting wire (coated
conductors).}
\end{figure}

Let us consider a superconductor containing inclusions of a normal
phase, which are out of contact with one another. These inclusions
may be formed by the fragments of different chemical composition,
as well as by the domains of reduced superconducting order
parameter. The columnar defects of such a kind can readily be
created in superconducting film at the sites of defects on the
boundary with the substrate (see Fig.~\ref{fig2}). A passage of
electric current through a superconductor is linked with the
vortex dynamics because the vortices are subjected to the Lorentz
force created by the current. In its turn, the motion of the
magnetic flux transferred by vortices induces an electric field
that leads to the energy losses. In HTS's the vortex motion is of
special importance because of large thermal fluctuations and small
pinning energies. \cite{blatter} Here we will consider the
simplified model of one-dimensional line pinning when a vortex
filament is trapped by the set of pinning centers. \cite{blatter2}
A superconductor containing isolated clusters of a normal phase
allows for effective pinning, because the vortices cannot leave
them without crossing the surrounding superconducting space. The
clusters present the sets of normal-phase inclusions, which are
united by the common trapped flux and surrounded by the
superconducting phase. The flux can be created both by an external
source (e.g., during magnetization in field-cooling regime) and by
the transport current (in self-field regime). The magnetic flux
remains to be trapped in the normal-phase clusters till the
Lorentz force created by the transport current exceeds the pinning
force.

\begin{figure}
\includegraphics{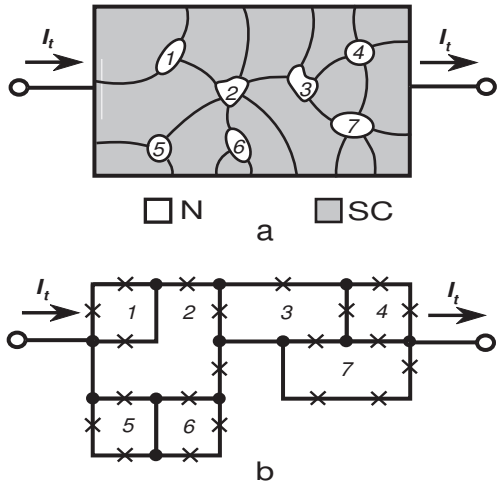}
\caption{\label{fig3} Equivalent weak-link-network circuit
representation of superconductor with normal-phase clusters. (a) -
the percolation chart (superconducting space is shaded, weak links
are shown by the curves joining together the normal-phase
clusters); (b) - the electric circuit (the superconducting loops
are shown by lines, the crosses on them denote weak links).}
\end{figure}

If the relative portion of superconducting phase exceeds the
percolation threshold, there is a superconducting percolation
cluster carrying a transport current. In such a structure magnetic
flux is locked in finite clusters of a normal phase (see
Fig.~\ref{fig3}(a)). If the transport current is passed through
the sample, the trapped flux remains unchanged as long as the
vortices are still held in the normal-phase clusters. The larger
clusters have to allow for a weaker pinning than smaller ones,
because the larger the cluster size, the more entry points into
weak links, through which the vortices can pass, are located along
its boundary. When the current is increased, the magnetic flux
will break away first from the clusters of smaller pinning force,
and therefore, of larger size. Magnetic flux trapped into a single
cluster is proportional to its area $A$. Therefore, the decrease
in the total trapped flux $\Delta \Phi $ is proportional to the
number of the clusters of area larger than a preset value $A$. So
that value can be expressed with the cumulative probability
function $W\left( A\right) =\Pr \left\{ \forall A_{j}<A\right\} $,
which is equal to the probability to find the cluster of area
$A_{j}$ smaller than an upper bound of $A$:
\begin{equation}
\frac{\Delta \Phi }{\Phi }=1-W\left( A\right)  \label{prob1}
\end{equation}

The left hand side of this formula is equal to the relative decrease in the
total trapped flux caused by the transport current of the same amplitude as
the depinning current of the cluster of area $A$, and the right hand side
gives the provability to find the cluster of area greater than $A$ in the
whole population.

When the current is increased, the magnetic flux starts to break
away from the clusters of pinning force weaker than the Lorentz
force created by the transport current. The vortices, driven by
this force, must cross the surrounding superconducting space, and
they will first do that through the weak links, connecting the
normal-phase clusters (see Fig.~\ref{fig3}(a) where weak links are
shown by the black curves connecting the white normal-phase
clusters). In such a system depinning has a percolative character,
\cite{yamafuji,ziese,ziese2} because unpinned vortices move
through the randomly generated channels created by weak links. In
Fig.~\ref {fig3}(b) the equivalent weak-link-network circuit
representation of such a system is shown. The normal-phase
clusters are presented by the cells of the net interlaced by the
superconducting loops. Each of the loops contains weak links
(shown by the crosses), which join the adjacent cells and so
enable the vortices to pass from one cluster to another.

Weak links form readily in HTS characterized by an extremely short
coherence length. \cite{blatter,scalapino,kerchner} Various
structural defects, which would simply cause some additional
scattering at long coherence length, give rise to the weak links
in HTS. On a mesoscopic scale twin boundaries are mainly
responsible for weak link existence. \cite
{pastoriza,kupfer,maggio,liu} Twins form especially readily in
YBCO superconductors inasmuch as their unit cell is only close to
the orthorhombic one. The twins can be spaced up to several
nanometers apart, so even single crystal may have the fine
substructure caused by twins. Magnetic and transport properties of
HTS depend strongly on the orientation of twin planes with respect
to the applied magnetic field. \cite{oussena,oussena2} The flux
can easily move along the weak links formed by twins.
\cite{duran,duran2,welp,wijn} At last, on a macroscopic scale
there are manifold structural defects which can form weak links:
that may be grain or crystallite boundaries as well as barriers
arising from the secondary degrading the non-stoichiometric
crystal into the domains with a high and low content of oxygen.
\cite{kerchner,hanisch,rykov,schuster,kilic} Moreover, a magnetic
field further reduces a coherence length, thus resulting in more
easy weak links formation. \cite{sonier} In conventional
low-temperature superconductors weak links can be formed due to
the proximity effect at the sites of minimum distance between the
next normal-phase clusters.

As soon as the transport current is turned on, this one is added
to all the persistent currents, maintaining the constant trapped
magnetic flux. Each of these currents is circulating through the
superconducting loop around the normal-phase cluster wherein the
corresponding portion of the magnetic flux is locked. The loop
contains weak links that join the adjacent normal-phase clusters.
As the transport current is increased, there will come a point
when the overall current flowing through the weak link will exceed
the critical value, so this link will switch into a resistive
state. During this process the space distribution of the currents
throughout the superconducting cluster is changed in such a way
that the resistive subcircuit will be shunted by the
superconducting paths where weak links are not damaged yet.
Magnetic field created by this re-distributed transport current
acts via the Lorentz force on the current circulating around the
normal-phase cluster. As a result, the magnetic flux trapped
therein will be forced out through the resistive weak link, which
has become permeable to the vortices.

Thus, whatever the microscopic nature of weak links may be, they
form the channels for vortex transport. It appears that according
to their configuration each normal-phase cluster has its own value
of the critical current, which contributes to the overall
statistical distribution. By the critical current of the cluster
we mean the current of depinning, namely, such a current at which
the magnetic flux ceases to be held inside the cluster of a normal
phase. When a transport current $I$ is gradually increased, the
vortices will break away first from clusters of small pinning
force, and therefore, of small critical current. Thus the decrease
in the trapped magnetic flux is proportional to the number of all
the normal-phase clusters of critical currents less than a preset
value. Therefore, the relative change in the trapped flux can be
expressed with the cumulative probability function $F\left(
I\right) =\Pr \left\{ \forall I_{j}<I\right\} $ for the
distribution of the critical currents of clusters:
\begin{equation}
\frac{\Delta \Phi }{\Phi }=F\left( I\right)  \label{probi2}
\end{equation}

The critical current distribution $F=F(I)$ is related to the cluster area
distribution $W=W(A)$, because the cluster of a larger size has more weak
links over its boundary with the surrounding superconducting space, and,
consequently, the smaller depinning current.

\section{GEOMETRIC PROBABILITY ANALYSIS OF THE DISTRIBUTION OF WEAK LINKS
OVER THE CLUSTER BOUNDARY}

In order to find out the relationship between the distribution of the
critical currents of the clusters and the distribution of their areas, the
distribution of entry points into weak links over the boundary of a
normal-phase cluster should be analyzed. We will examine the geometric
properties of the normal-phase clusters in its cross-section by the plane
where the transport current flows. The vortices are moving transversally to
this plane so in order to leave the normal-phase cluster they have to get
over the barrier at the boundary of that cross-section. We will also suppose
that the boundaries of the cluster cross-section are statistically
self-similar.

The problem of exit of a vortex from a normal cluster represents
the two-dimensional analogue of a problem of a random walk
particle reaching a border. \cite{spitzer,gardiner} At the same
time, unlike the classic problem on the distribution of the exit
points, here the boundary of the area is not absorbing all over.
In order to leave the cluster the vortex has to enter into one of
the weak links, which are randomly arranged along the cluster
perimeter. There are only discrete absorption points, which are
located just at the sites where weak links are going out on the
cluster boundary. In other words, these points are the points of
the entry of vortices into weak links, or simply, entry points.
For simplicity it will have been supposed that after the vortex
reaches the entry point, it passes all the way between two
adjacent normal-phase clusters without being trapped inside the
weak link itself. Here the magnetic flux is transferred by
Josephson vortices. The Josephson penetration depth is large
enough in the considered materials, so the size of the region,
where the vortex is localized, much exceeds the characteristic
length of all possible structural defects that can occur along the
transport channel. Thus the probability that such a vortex, driven
by the Lorentz force, will be trapped in passing through a weak
link is very small. This assumption agrees well with the results
of research on the magnetic flux motion along weak links
\cite{wijn,dorog,fangohr} including twins.
\cite{duran,duran2,welp,turch,welp2} At the same time, it allows
us to highlight the role played by the cluster boundary in the
magnetic flux dynamics.

Let us consider the distribution of entry points over perimeter of a
normal-phase cluster. Generally, this distribution varies from one cluster
to another, so that each normal-phase cluster has the entry point
distribution function $\psi \left( l\right) $ of its own, which belongs to
some function class $\Omega $. Here $l$ is the co-ordinate measured along
the cluster perimeter, so $l\in \left( 0,\text{ }P\right) .$ \ In this
context the functions of class $\Omega $ are random elements of the
statistical distribution. The probability distribution of functions $\psi
\left( l\right) $ over all the clusters can be characterized by the
functional Pr$\{\psi \left( l\right) \}$, which is equal to the probability
of finding a given function $\psi \left( l\right) $.

In the most general way the geometric probability analysis of the
entry point into weak link distribution can be carried out by
means of path integral technique. \cite{feynman} The probability
that there is the function $\psi \left( l\right) $ in class
$\Omega $ may be expressed by the path integral
\[
\Pr \left\{ \psi \left( l\right) \right| \Omega \}=\int\limits_{(\Omega
)}D\psi \left( l\right) \Pr \left\{ \psi \left( l\right) \right\}
\]

Therefore, the most probable function of entry point distribution is the
mean over all functions of class $\Omega $%
\begin{equation}
\Psi \left( l\right) \equiv \overline{\psi \left( l\right) }%
=\int\limits_{(\Omega )}D\psi \left( l\right) \psi \left( l\right) \Pr
\left\{ \psi \left( l\right) \right\}  \label{path3}
\end{equation}

The path integral Fourier transform on the probability functional Pr$\{\psi
\left( l\right) \}$ represents the characteristic functional
\begin{equation}
H\left[ k(l)\right] =\frac{\int\limits_{\left( \Omega \right) }{\cal D}\psi
\left( l\right) \,\exp \left( i\oint dl\,k\left( l\right) \psi \left(
l\right) \right) \Pr \left\{ \psi \left( l\right) \right\} }{%
\int\limits_{\left( \Omega \right) }{\cal D}\psi \left( l\right) \,\Pr
\left\{ \psi \left( l\right) \right\} }  \label{fourier4}
\end{equation}
where $k=k\left( l\right) $ are the functions of a reciprocal function set,
and integration in the kernel $\exp \left( i\oint dl\,k\left( l\right) \psi
\left( l\right) \right) $ is carried out over the cluster perimeter.

The characteristic functional is the path integration analog for the usual
moment-generating function. The probability functional $\Pr \left\{ \psi
\left( l\right) \right\} $ can be written as an inverse path integral
Fourier transform on the characteristic functional
\[
\Pr \left\{ \psi \left( l\right) \right\} =\int Dk\left( l\right) \exp
\left( -i\oint dl\,k\left( l\right) \psi \left( l\right) \right) H\left[
k\left( l\right) \right]
\]
where the path integration is carried out on the reciprocal function space.

In the simplest case, when all the clusters are of equal entry point
distribution, which coincides with the most probable one of Eq.~(\ref{path3}%
), the probability functional $\Pr \left\{ \psi \left( l\right) \right\} $
is zero for all $\psi \left( l\right) $ that differ from $\Psi \left(
l\right) $, whereas $\Pr \left\{ \Psi \left( l\right) \right\} =1$.
Therefore, the characteristic functional of Eq.~(\ref{fourier4}) becomes

\begin{equation}
H\left[ k\left( l\right) \right] =\exp \left( i\oint dlk\left( l\right) \Psi
\left( l\right) \right)  \label{char5}
\end{equation}

If all entry points had fixed co-ordinates $l_{j}$ instead of the random
ones, their distribution would be $\psi \left( l\right) =\beta
\mathop{\textstyle\sum}%
\nolimits_{j=1}^{N}\delta \left( l-l_{j}\right) $ , where $N$ is the number
of entry points along the cluster perimeter, $\delta \left( l\right) $ is
Dirac delta function. The constant $\beta $ is being chosen to normalize the
distribution function $\psi \left( l\right) $ to unity, so that $\beta N=1$.

Now suppose that all the points of entries into weak links are randomly
distributed with uniform probability over the cluster perimeter, so the
probability to find any $j$-th point within some interval $dl_{j}$ is
proportional to its length. In that case the characteristic functional of
Eq.~(\ref{fourier4}) takes the form
\begin{eqnarray}
H\left[ k\left( l\right) \right]  &=&\frac{\oint
\mathop{\textstyle\prod}%
\limits_{j=1}^{N}dl_{j}\,\exp \left( i\beta
\sum\limits_{j=1}^{N}\oint dl\,k\left( l\right) \delta \left(
l-l_{j}\right) \right) }{\oint
\prod\limits_{j=1}^{N}dl_{j}}  \nonumber \\
&=&\frac{1}{P^{N}}\oint
\mathop{\textstyle\prod}%
\limits_{j=1}^{N}dl_{j}\,\exp \left( i\beta
\sum\limits_{j=1}^{N}k\left(
l_{j}\right) \right)   \nonumber \\
&=&\frac{1}{P^{N}}%
\mathop{\textstyle\prod}%
\limits_{j=1}^{N}%
\displaystyle\oint %
dl_{j}\,e^{i\beta k\left( l_{j}\right) }  \label{cf6}
\end{eqnarray}

Expanding the function $e^{i\beta k\left( l\right) }$ in a power series at $%
N>>1$, and taking into account the condition $\beta N=1$, we may write
\[
\frac{1}{P}\oint dle^{i\beta k\left( l\right) }=\exp \left( i\frac{\beta }{P}%
\oint dl\,k\left( l\right) \right)
\]
that, after substitution into Eq.~(\ref{cf6}), gives
\begin{equation}
H\left[ k\left( l\right) \right] =\exp \left( i\frac{\beta N}{P}\oint
dl\,k\left( l\right) \right)  \label{charf7}
\end{equation}

The found characteristic functional of Eq.~(\ref{charf7}) has the form of
Eq.~(\ref{char5}) for the function of uniform distribution of entry points
\begin{equation}
\Psi \left( l\right) =\frac{1}{P}  \label{uni8}
\end{equation}
\qquad \qquad This means that all the clusters have the same uniform
distribution of the entry points of Eq.~(\ref{uni8}), for which the
probability of finding a weak link at any point of the perimeter is
independent of its position.

Let us suppose that concentration of entry points into weak links per unit
perimeter length $n=\overline{N}/P$ is constant for all clusters, and all
the clusters are statistically self-similar. In this case the mean number of
entry points $\overline{N}$ along the cluster perimeter is proportional to
its length:
\begin{equation}
\overline{N}=\oint n\left( l\right) dl=nP  \label{mean9}
\end{equation}

Next step will consist in finding the relationship between the
size of a cluster and its critical current. The pinning force
corresponds to such a current at which the vortices start to break
away from the cluster. As the transport current is increasing, the
Lorentz force, which expels the magnetic flux, increases as well.
The vortices start to leave the normal-phase cluster when the
Lorentz force becomes greater than the pinning force. At the same
time, growing in current will result in re-distribution of the
magnetic flux, which will penetrate deeper and deeper into a
transition layer on that side of the surrounding superconducting
space where the Lorentz force is directed (see
Fig.~\ref{fig2}(b)). In order to leave the normal-phase cluster,
vortices have to reach the entry points into weak links. The exit
of the magnetic flux can be considered as the result of random
walks of vortices driven by the Lorentz force, which is pushing
them into weak links. The similar approach has been successfully
applied in analyzing the magnetic flux penetration in SQUID
arrays. \cite{dorog} The mean number of the entry points
$\overline{N}$ available on the cluster perimeter gives the
probability measure of the number of the random walk outcomes,
which are favorable for the vortex to go out. In the case of the
uniform entry point distribution, from Eq.~(\ref{mean9}) it follows that $%
\overline{N}\propto P$, so the perimeter length also represents the
probability measure of the amount of favorable outcomes for vortex to leave
the cluster. The more entry points into weak links are accessible for random
walk vortices, the smaller is the Lorentz force required to push the flux
out. Hence, we may write the following relation between the critical current
of the cluster and its geometric size:
\begin{equation}
I\propto \frac{1}{\overline{N}}\propto \frac{1}{P}  \label{geocur10}
\end{equation}

This expression is true for the simplest case of uniform distribution of
entry points, which is assumed to be the same for all clusters. Such a
simplification allows us to emphasize that in the case being considered the
magnetic flux is held in the normal-phase cluster by its boundary.

Thus, to deal with the distribution function of Eq.~(\ref{prob1}),
the relation between perimeter and area of clusters should be
studied. It might be natural to suppose that the perimeter-area
relation obeys the well known geometric formula: $P\propto
\sqrt{A}$. However, it would be a very rough approximation,
because this relation holds for Euclidean geometric objects only.
As was first found in Ref.~\onlinecite{pla2000}, the boundaries of
normal-phase clusters can be fractal, i.e. can have non-Euclidean
features. The fractal nature of such clusters exerts an
appreciable effect on the magnetic flux dynamics in
superconductors. \cite{prb,pla2001,pss}

\section{A Brief Introduction to Fractal Geometry}

The notion of a fractal as an object of fractional dimension was
first introduced by Mandelbrot \cite{mandelbrot} in 1967 and since
then it has received a lot of applications in various domains of
sciences. \cite{mandelbrot2,mandelbrot3,mandelbrot4,feder} This
concept is closely connected to ideas of scaling and
self-similarity. Self-similarity is invariance with respect to
scaling; in other words, an invariance relative to multiplicative
changes of scale. Whereas a usual periodicity is invariance with
respect to additive translations, self-similarity is a periodicity
in a logarithmic scale.

The simplest examples of self-similar objects are Cantor sets.
\cite {mandelbrot2} Such a set has measure zero and, at the same
time, has so many elements that it is uncountable. The Cantor set
may be constructed by such a way. Let us draw a line segment, and
erase its middle third. Then eliminate middle third of each
remaining part, and so on. The resulting set obtained by endless
erasing of the middle thirds of remaining intervals is called
``Cantor dust``. \cite{mandelbrot2} The Cantor dust forms a
self-similar set because it is invariant to scaling by a factor
$s=3$. But this set is not only a self-similar one, but it is a
fractal as well. Let us see what part of the set falls within one
$s$-th part of the original set, where $s$ is a scaling factor,
and ask what fraction of the set falls into that portion? In other
words, how many subsets, each is similar to the original set, are
there if the length is subdivided into $s$ parts? This number of
such subsets is equal to $N=2$, so one-half of the original set
falls into one third of the initial length. The fractal dimension
is defined as the logarithm of the subset number divided by the
logarithm of the scaling factor: $D=\ln N/\ln s=\ln 2/\ln
3=0.631$. This formula represents the relation between the subset
number and the scaling factor: $N=s^{D}$. Thus the fractal
dimension of the Cantor dust is less than unity. This fact
reflects its dust-like consistency compared to a usual line. An
Euclidean line is such a set that if we change the length scale,
we recover the same set of points. Hence the fractal dimension of
a line coincides with the topological dimension, and both of them
are equal to unity. Generally, a fractal object has a fractional
dimension. A fractal set is such a set for which the fractal
dimension strictly exceeds its topological dimension. \cite
{mandelbrot3}

The Cantor dust gives an example of determinate fractal that is
self-similar by deterministic construction. But the same approach
can be applied to the stochastic fractals, which are statistically
self-similar. \cite{mandelbrot4} The fundamental feature of any
fractals, both determinate and the stochastic ones, is that its
characteristic measures obey a scaling law that includes an
exponent called the fractal dimension.

Fractal dimension of statistically self-similar curve can be estimated by
covering the curve by square grid. The size of the grid cell $L\times L$
should be small enough for measuring all the bend of the curve. If the whole
curve (the boundary of the normal-phase cluster cross-section) goes in
square of side $L_{\max }$, then each of $N(L)$ self-similar subsets of that
curve will go in square of side $L=L_{\max }/s$, where $s$ is scaling
factor. Let us suppose that the curve is self-similar at any scaling factor.
The minimum number of such squares of side $L$ needed to cover the explored
curve is equal to
\[
N\left( L\right) =s^{D}=\left( \frac{L_{\max }}{L}\right) ^{D}\propto L^{-D}
\]
where $D$ is the fractal dimension of the curve (so-called
coastline dimension. \cite{mandelbrot2} So we have the formula
that determines the fractal dimension
\begin{equation}
D=\frac{\ln N\left( L\right) }{\ln \left( L_{\max }/L\right) }  \label{dim11}
\end{equation}
It is obvious that the magnitude of $D$ is always below the topological
dimension of a plane, which is equal to two.

The ratio between the size of measuring cell $L$ and the big square side $%
L_{\max }$ sets the unit of measurement of length. At $L_{\max }=1$ the
expression for the fractal dimension of Eq.~(\ref{dim11}) takes the form $%
D=-ln(N(L))/ln(L)$.

Thus, in order to find the fractal dimension it is necessary to count the
total number $N(L)$ of non-empty cells of area $L^{2}$, which cover the
curve. To attain the highest precision of estimate we have to get wide range
of $L$ and to average obtained values of $D$ over all the configurations.

The length of topologically one-dimensional fractal curve of size $L_{\max }$%
, measured with an accuracy of $L$, is equal to a product of the number of
cells, which cover that curve, and the linear size of the measuring cell
\begin{equation}
P=NL=\left( \frac{L_{\max }}{L}\right) ^{D}L=\frac{L_{\max }^{D}}{L^{D-1}}
\label{length12}
\end{equation}
In other words, if the whole curve goes in the scale of length $L_{\max }$,
then in the smaller scale of $L=L_{\max }/s$ (i. e. for the yardstick of a
smaller size) the same curve will consist of $N=s^{D}$ pieces of length $L$.
As is seen from the formula of Eq.~(\ref{length12}), the length of a fractal
curve is not well defined, because its value depends on the accuracy of the
measurements, and even diverges as the yardstick size $L$ is reduced
infinitely. At the same time, the area restricted by any closed (including
fractal) curve is well-defined finite quantity, which is proportional to its
squared linear size: $A\propto L_{\max }^{2}$. According to expression of
Eq.~(\ref{length12}), the length of the perimeter of the figure, formed by
the closed fractal curve, is proportional to its linear size raised to a
power of fractal dimension: $P\propto L_{\max }^{D}$. Hence it follows that
the perimeter of fractal figure and enclosed area have to obey a scaling law

\begin{equation}
P^{1/D}\propto A^{1/2}  \label{scaling13}
\end{equation}
The perimeter-area scaling law of Eq.~(\ref{scaling13}) expresses
the generalized Euclid theorem about measures of similar figures,
stating that the ratios of corresponding measures are equal when
reduced to the same dimension. \cite{mandelbrot3} This assertion
is valid both for Euclidean figures and for the fractal ones.

The fractal approach has been found to be most useful in an
investigation of inhomogeneous materials. There are many
possibilities both the determinate fractals and the stochastic
ones to be formed in composite superconductors. As an example of
the first kind the multi-layered structures prepared by
electron-beam deposition of superconductor (Nb) and normal metal
(Cu) layers with fractal stacking sequence on sapphire substrates
can be mentioned. \cite {sidorenko} In order to obtain stochastic
fractal clusters, it is essential that the process like the
diffusion-limited aggregation would take place in the course of
the synthesis of material. \cite{feder} The similar process can be
realized, for instance, when thin films are evaporated. So, in
Ref.~ \onlinecite{laibowitz} the films of fractal structure has
been grown by vapor deposition of gold on silicon substrate with
silicon nitride buffer sub-layer. It is worthy of note that
porous, random, or highly ramified clusters do not necessarily all
are fractals. A fractal cluster has such a property that its
characteristic measures (in what follows - the perimeter and the
enclosed area) have to obey the certain scaling law that includes
an exponent named fractal dimension. \cite{mandelbrot2,feder}

The fractal clusters can be also formed in such highly
inhomogeneous materials as high-temperature superconductors. So,
the fractal dissipative regime has been observed in
high-resolution measurements of dynamical resistance of BSCCO and
BPSCCO composites containing normal-phase inclusions of Ag.
\cite{prester} The fractal properties of the normal-phase clusters
contained in YBCO films, which were prepared by magnetron
sputtering on sapphire substrate with a cerium oxide buffer
sub-layer, have been found in Ref.~\onlinecite{pla2000}.
Percolation clusters are another example of fractals in
superconductors. \cite{alexander,hong} Although, mathematically,
the percolation cluster is a fractal at the threshold point only,
the fractal approach works well for any clusters which have a
scaling feature. \cite{pike,gefen,havlin} In that case the
normal-phase clusters may be formed by inclusions of different
chemical composition, as well as the domains of reduced
superconducting order parameter can act as such clusters. The
existence of the fractal inclusions of this kind can be
demonstrated by the fractal dissipation, which has been observed
in non-textured polycrystalline YBCO and GdBCO bulk samples.
\cite{prester} The fractal structure of clusters near the
percolation threshold in epitaxial YBCO films has been fully
considered in Ref.~\onlinecite{baziljevich}. Of special interest
are the works of Refs.~\onlinecite{surdeanu,surdeanuPhC}, where
the fractal penetration of magnetic flux in thin HTS films has
been investigated by magneto-optical technique. Epitaxial TlBCO
films were grown by magnetron sputtering on SrTiO$_{3}$
substrates, and YBCO films were prepared by pulsed laser
deposition on NdGaO$_{3}$ substrates. The cluster structure of
such films is clearly visible on the Atomic Force Microscopy
picture published in Ref.~\onlinecite{surdeanu}, whereas the
magnetic flux, penetrating into the sample from the outside, has
well-definite fractal front.

\section{Fractal Geometry of Normal-Phase Clusters}

The scaling law of Eq.~(\ref{scaling13}) furnishes a clue to find
relation between the critical current of the cluster and its
geometric size. Using the formula of Eq.~(\ref{geocur10}), which
links the cluster size and its current of depinning, can we get
the following expression: $I=\alpha A^{-D/2} $, where $\alpha $ is
the form factor, and $D$ is the fractal dimension of the cluster
boundary. In the general way the cluster area distribution can be
described by gamma distribution, which is appropriate for the most
part of practically realizable cases.
\cite{pla2001,pss,tpl2002,pla2002} This distribution is
characterized by the following cumulative probability function:
\begin{equation}
W\left( A\right) =\left( \Gamma \left( g+1\right) \right) ^{-1}\gamma \left(
g+1,\frac{A}{A_{0}}\right)  \label{cuma14}
\end{equation}
where $\Gamma \left( \nu \right) $ is Euler gamma function, $\gamma \left(
\nu \right) $ is the incomplete gamma function, $A_{0}$ and $g$ are the
parameters of gamma distribution that control the mean area of the cluster $%
\overline{A}=\left( g+1\right) A_{0}$ and its variance $\sigma
_{A}^{2}=\left( g+1\right) A_{0}^{2}$.

In accordance with starting formulas of Eq.~(\ref{prob1}) and Eq.~(\ref
{probi2}), gamma distribution of cluster areas of Eq.~(\ref{cuma14}) gives
rise to the critical current distribution of the form:
\begin{equation}
F\left( i\right) =\left( \Gamma \left( g+1\right) \right) ^{-1}\Gamma \left(
g+1,Gi^{-2/D}\right)  \label{cumi15}
\end{equation}
where $G\equiv \left( \theta ^{\theta }/\left( \theta
^{g+1}-\left( D/2\right) \exp \left( \theta \right) \Gamma \left(
g+1,\theta \right) \right) \right) ^{2/D}$, $\theta \equiv
g+1+D/2$,\ $\Gamma \left( \nu ,z\right) $ is the complementary
incomplete gamma function, $i=I/I_{c}$ is the dimensionless
electric current, $I_{c}=\alpha \left( A_{0}G\right)^{-D/2}$ is
the critical current of the transition into a resistive state. The
found cumulative probability function of Eq.~(\ref{cumi15}) allows
to derive the probability density $f(i)\equiv dF/di$ for the
critical current distribution:
\begin{equation}
f(i)=\frac{2G^{g+1}}{D\Gamma (g+1)}i^{-\left( 2/D\right) (g+1)-1}\exp \left(
-Gi^{-2/D}\right)  \label{densi16}
\end{equation}
This distribution allows us to fully describe the effect of the transport
current on the trapped magnetic flux taking into account the fractal
properties of the normal-phase clusters. Let us note that the probability
density for the critical current distribution of Eq.~(\ref{densi16}) is
equal to zero at $i=0$, which implies the absence of any contribution from
negative and zero currents. This will allow us to avoid any artificial
assumption about the existence of a vortex liquid, having finite resistance
in the absence of transport currents due to the presence of free vortices: $%
r\left( i\rightarrow 0\right) \neq 0$. Such an assumption is made,
for example, in the case of normal distribution of critical
currents. \cite{brown}

\begin{figure}
\includegraphics{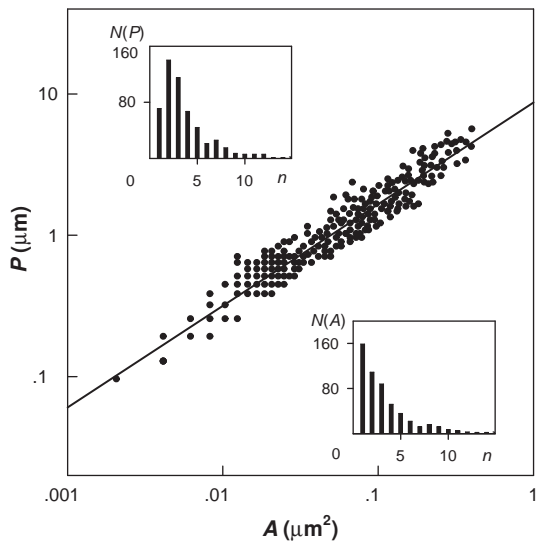}
\caption{\label{fig7} Perimeter-area relationship for the
normal-phase clusters with fractal boundary. The chart is built
for the primary sampling containing 528 normal-phase clusters. The
insets represent histograms of sampling for corresponding
distributions.}
\end{figure}

In order to clear up how the developed approach can be used in
practice, the geometric probability analysis of electron
photomicrographs of superconducting films was carried out. For
this purpose electron photomicrographs of YBCO film prepared by
magnetron sputtering on sapphire substrate with a cerium oxide
buffer sub-layer have been scanned. The normal phase has occupied
20\% of the total surface only, so the transport current can flow
through the sufficiently dense percolation superconducting
cluster. The perimeters and areas of clusters have been measured
by covering their digitized pictures with a square grid of spacing
60$\times $60\thinspace nm$^{2}$. The results of the statistical
treatment of these data are presented in Fig.~\ref {fig7}. The
primary sampling has contained 528 normal-phase clusters
located on the scanned region of a total area of 200\thinspace $\mu $m$^{2}$%
. The distribution of the cluster areas is fitted well to
exponential cumulative probability function, as is shown on the
histogram in the lower inset of Fig.~\ref{fig7}. The number of
clusters that fall within the assigned rank is plotted on the
ordinate of this graph; the rank number is plotted on the
abscissa. A high skewness (1.765) as well as the statistically
insignificant (5\%) difference between the sample mean area of the
cluster and the standard deviation also attests that there an
exponential distribution of the cluster areas holds true. This
distribution is a special case of gamma distribution for which
$g=0$, so the cumulative probability function of
Eq.~(\ref{cuma14}) can be simplified to the following form:
\begin{equation}
W\left( A\right) =1-\exp \left( -\frac{A}{\overline{A}}\right)
\label{probfunc17}
\end{equation}
The exponential distribution has only one characteristic parameter - the
mean cluster area $\overline{A}$.

The obtained data allow us to find the perimeter-area relation for
the normal-phase clusters. All the points in Fig.~\ref{fig7} fall
on a straight line in double logarithmic scale with correlation
coefficient of 0.929. Accordingly to the scaling relation of
Eq.~(\ref{scaling13}), the slope of the regression line gives the
estimate of fractal dimension of the cluster perimeter, which is
equal to $D=1.44\pm 0.02$. The graph in Fig.~\ref {fig7} shows
that the scaling law for perimeter and area, which is inherent to
fractals, is valid in the range of almost three orders of
magnitude in cluster area. The perimeter-area scaling behavior
means that there is no characteristic length scale in the range of
two orders of linear size of the normal-phase cluster. Whatever
the shape and size of the clusters may be, all the points fall
closely on the same straight line in logarithmic scale; so that
there are no apparent kinks or bends on the graph. This point that
the found value of the fractal dimension differs appreciably from
unity engages a great attention. What this means is that the
fractal properties of the cluster boundary are of prime importance
here.

\begin{figure}
\includegraphics{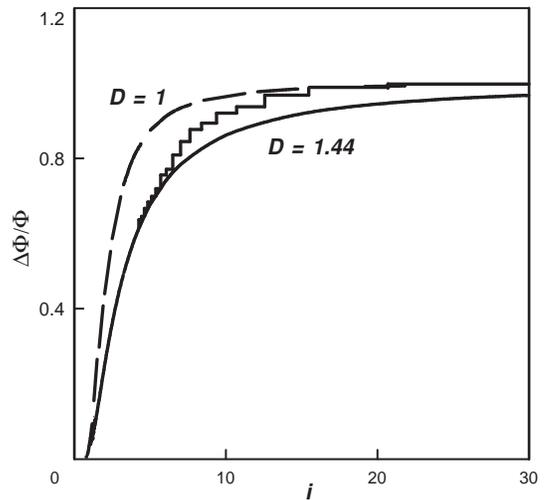}
\caption{\label{fig8} Crossover between the fractal and Euclidean
regimes on dependence of the magnetic flux trapped in fractal
clusters of a normal phase on a transport current. The solid line
shows the decrease in trapped flux for the fractal clusters of
boundary dimension $D=1.44$; the dotted line corresponds to the
case of Euclidean clusters ($D=1$); step line is the sample
empirical function of critical current distribution.}
\end{figure}

The geometric probability properties of the normal-phase clusters are
responsible for main features of the critical current statistical
distribution. Now, knowing the fractal dimension of the cluster boundaries,
the change in the trapped magnetic flux caused by the transport current can
be found with aid of Eq.~(\ref{probi2}). The exponential distribution of the
cluster areas of Eq.~(\ref{probfunc17}) gives rise to the
exponential-hyperbolic distribution of critical currents
\begin{equation}
F\left( i\right) =\exp \left( -\left( \frac{2+D}{2}\right)
^{2/D+1}i^{-2/D}\right)  \label{exhyp18}
\end{equation}
which follows from Eq.~(\ref{cumi15}) at $g=0$. The effect of a
transport current on the trapped magnetic flux is illustrated in
Fig.~\ref{fig8} for the case of Euclidean clusters (dotted curve),
and for the clusters of found fractal dimension $D=1.44$ (solid
curve).

In order to get the relationship between the dynamics of the trapped
magnetic flux and the geometric properties of the superconducting structure
the empirical function of the distribution of the critical currents $F^{\ast
}=F^{\ast }(i)$ has been found. First, the empirical distribution function $%
W^{\ast }=W^{\ast }(A)$ for the sampling of the areas of the normal-phase
clusters has been obtained. The value of $W^{\ast }(A)$ was calculated for
each order statistic as the relative number of clusters of area smaller than
a given value $A$. Next, the empirical distribution of the critical currents
was computed for the same order statistics using the formulas:
\[
\left.
\begin{array}{c}
F^{\ast }=1-W^{\ast } \\
i=\left( \left( 2+D\right) /2\right) ^{\left( 2+D\right) /2}\left( \overline{%
A}/A\right) ^{D/2}
\end{array}
\right\}
\]

This function, shown in Fig.~\ref{fig8} by the step line, gives a
statistical estimate of the cumulative probability function of
Eq.~(\ref {exhyp18}). As is seen from this figure, both
distributions coincide well in the range of currents $i<6$.
Starting with the value of the current $i=6$ the crossover from
the fractal regime to the Euclidean one is observed. This
transition into the Euclidean regime is over at large transport
currents, when the magnetic flux changes mainly for the breaking
of the vortices away from the small clusters (as the smaller
clusters have the larger pinning force). The observed crossover
has its origin in the finite resolution capability of measuring
the cluster sizes. When estimating the fractal dimension, we have
to take into account that the resolution of the measurement of any
geometric sizes is finite. The peculiarity of the topologically
one-dimensional fractal curve is that its measured length depends
on the measurement accuracy. \cite{mandelbrot2} In our case such a
fractal curve is represented by the boundary of the normal-phase
cluster. That is why just the statistical distribution of the
cluster areas, rather than their perimeters, is fundamental for
finding the critical current distribution. The topological
dimension of perimeter is equal to unity and does not coincide
with its fractal dimension, which strictly exceeds the unity.
Therefore the perimeter length of a fractal cluster is not well
defined, because its value depends on the yardstick size. On the
other hand, the topological dimension of the cluster area is the
same as the fractal one (both are equal to two). Thus, the area
restricted by the fractal curve is a well-defined quantity.

Taking into account the finite resolution effect, the perimeter-area
relationship of Eq.~(\ref{scaling13}) can be re-written in the following
form
\begin{equation}
P(\delta )\propto \delta ^{1-D}\left( A\left( \delta \right) \right) ^{D/2}
\label{resol19}
\end{equation}
where $\delta $ is the yardstick size used to measure this length, $(1-D)$
is the Hausdorff codimension for the Euclidean one-dimensional space. This
relation holds true when the yardstick length is small enough to measure
accurately the boundary of all smallest clusters in sampling. When the
resolution is deficient, the Euclidean part of the perimeter length will
dominate the fractal one, so there is no way to find the fractal dimension
using the scaling relation of Eq.~(\ref{resol19}). It means that if the
length of a fractal curve was measured too roughly with the very large
yardstick, its fractal properties could not be detected, and therefore such
a geometrical object would be manifested itself as Euclidean one. It is just
the resolution deficiency occurs at the crossover point in Fig.~\ref{fig8}%
. Starting with the cluster area less than some value (which
corresponds to the currents of $i>6$) it is impossible to measure
all ``skerries'' and ``fjords'' on the cluster coastlines, whereas
all the clusters of area less than the size of the measuring cell
(3600\thinspace nm$^{2}$ that relates to the currents of $i>23$),
exhibit themselves as objects of Euclidean boundaries. This
resolution deficiency can be also observed in Fig.~\ref {fig7}:
some points at its lower left corner are arranged discretely with
the spacing equal to the limit of resolution (60\thinspace nm),
because some marks for smallest clusters coincide for the finite
resolution of the picture digitization procedure.

\begin{figure}
\includegraphics{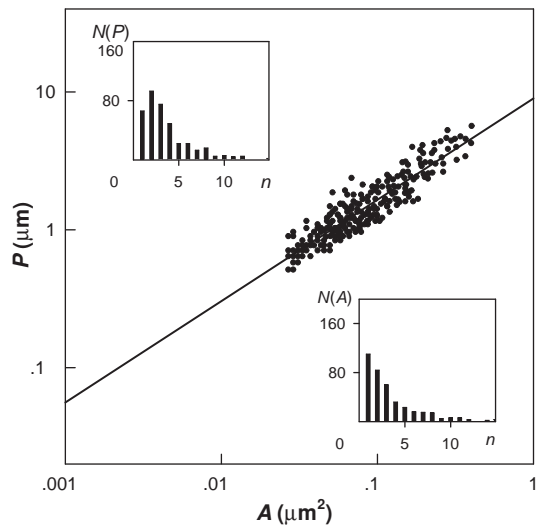}
\caption{\label{fig9} Perimeter-area relationship for the
normal-phase clusters with fractal boundary. The chart is built
for the truncated sampling containing 380 normal-phase clusters.
The insets represent histograms of sampling for corresponding
distributions.}
\end{figure}

The fractal dimension was found above by means of regression
analysis of the whole primary sampling, where the very small
clusters of sizes lying at the breaking point of the resolution
limit were also included. In order to evaluate how the finite
resolution affects the accuracy of the estimation of the fractal
dimension, all the points, for which the resolution deficiency was
observed, were eliminated from the primary sampling. So the
truncated sampling has been formed in such a way that only 380
clusters, for which the resolution deficiency is not appeared yet,
have been selected from the primary sampling. The regression
analysis of the truncated sampling is presented in
Fig.~\ref{fig9}. The least squares treatment of
perimeter-area data gives the adjusted magnitude of the fractal dimension: $%
D=1.47\pm 0.03$. The found value virtually does not differ from the previous
one within the accuracy of the statistical estimation, whereas the
correlation coefficient (which becomes equal to 0.869) falls by six
hundredth only. Therefore, we can conclude that the found estimate of the
fractal dimension is robust.

\section{THE PINNING GAIN FOR THE MAGNETIC FLUX TRAPPED IN FRACTAL CLUSTERS
OF A NORMAL PHASE}

\begin{figure}
\includegraphics{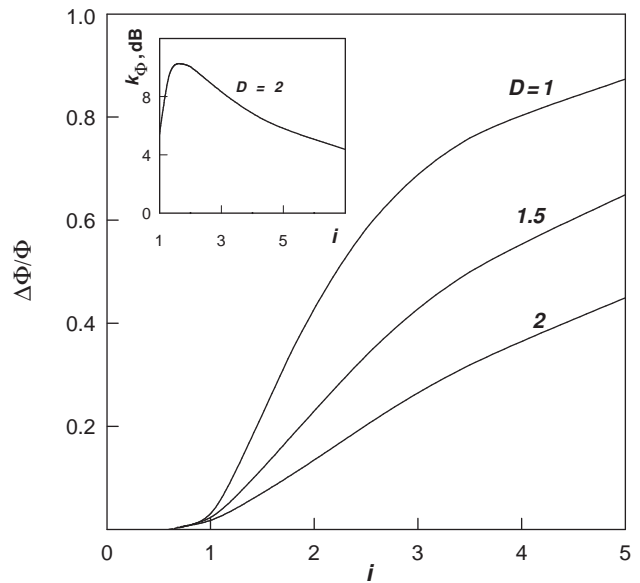}
\caption{\label{fig10} Effect of a transport current on the
magnetic flux trapped in fractal clusters of a normal phase. The
inset shows the pinning gain for the different fractal dimension
of the cluster boundary.}
\end{figure}

It has been revealed that the fractality of the cluster boundaries
intensifies the pinning. As can be seen from Fig.~\ref{fig10}, the
decrease in the trapped magnetic flux at the same value of the
transport current is less for larger fractal dimension. The
relative change in the trapped flux $\Delta \Phi /\Phi $, which,
according to formula of Eq.~(\ref {probi2}), can be calculated
from Eq.~(\ref{exhyp18}), defines the density of vortices n broken
away from the pinning centers by the current $i$:
\begin{equation}
n\left( i\right) =\frac{B}{\Phi _{0}}\int_{0}^{i}f\left( i^{\prime }\right)
di^{\prime }=\frac{B}{\Phi _{0}}\frac{\Delta \Phi }{\Phi }  \label{vort20}
\end{equation}
where $B$ is the magnetic field, $\Phi _{0}\equiv hc/\left(
2e\right) $ is the magnetic flux quantum ($h$ is Planck's
constant, $c$ is the velocity of light, $e$ is the electron
charge). The graph of the change in the trapped flux as a function
of transport current, shown in Fig.~\ref{fig10}, coincides
qualitatively with the curves obtained in experiments on the
magnetization of YBCO films subjected to current pulses.
\cite{tpl99,pss99} Figure \ref{fig10} also displays such an
important property of superconducting structure containing fractal
clusters of a normal phase: the fractality intensifies the
magnetic flux trapping, hindering its breaking away from pinning
centers, and thereby enhances the critical current which sample is
capable to withstand, remaining in a superconducting state. The
pinning amplification can be characterized by the pinning gain
factor
\begin{equation}
k_{\Phi }\equiv 20\log _{10}\frac{\Delta \Phi \left( D=1\right) }{\Delta
\Phi \left( \text{current value of }D\right) }\text{ , \ \ \ \ \ dB}
\label{pingain21}
\end{equation}
which is equal to relative decrease in the fraction of magnetic
flux broken away from fractal clusters of fractal dimension $D$
compared to the case of Euclidean ones. Dependencies of the
pinning gain on the transport current for different fractal
dimension at $g=0$ are shown in the inset of Fig.~\ref {fig10}.
The highest amplification (about 10\thinspace dB) is reached when
the cluster boundaries have the greatest possible fractality.
Figure \ref{fig10} demonstrates that with increase in fractal
dimension the trapped magnetic flux is changed less and less by
the action of the transport current. The pinning gain of
Eq.~(\ref{pingain21}) characterizes the properties of a
superconductor in the range of the transport currents
corresponding to a resistive state ($i>1$). At smaller current the
total trapped flux remains nearly unchanged (see Fig.~\ref{fig10})
for lack of pinning centers of such small critical currents, so
the breaking of the vortices away has not started yet. When the
vortices start to leave the normal-phase clusters and move through
the weak links, their motion induces an electric field, which, in
turn, creates the voltage drop across the sample. Therefore, the
passage of electric current is accompanied by the energy
dissipation. As for any hard superconductor (that is to say,
type-II, with pinning centers) this dissipation does not mean the
destruction of phase coherence yet. Some dissipation always
accompanies any motion of a magnetic flux that can happen in a
hard superconductor even at low transport current. Therefore the
critical current in such materials cannot be specified as the
greatest non-dissipative current. Superconducting state collapses
only when a growth of dissipation becomes avalanche-like as a
result of thermo-magnetic instability.

\begin{figure}
\includegraphics{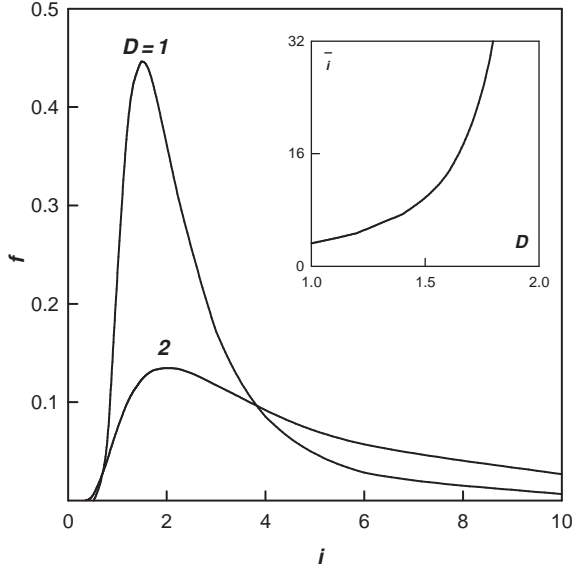}
\caption{\label{fig11} Influence of the fractal dimension of the
boundary of the normal-phase clusters on the critical current
distribution. The inset shows the dependencies of the mean
critical current on the fractal dimension.}
\end{figure}

The principal reason of pinning enhancement due to the fractality
of the normal-phase clusters lies in the fundamental properties of
the critical current distribution. Figure \ref{fig11} demonstrates
the peculiarities of the fractal probability density specified by
Eq.~(\ref{exhyp18}). As may be seen from this graph, the
bell-shaped curve of the distribution broadens out, moving towards
greater magnitudes of current as the fractal dimension increases.
It means that more and more of the small clusters, which can trap
the vortices best, are being involved in the game. Hence the
number of vortices broken away from pinning centers by the Lorentz
force is reducing, so the smaller part of a magnetic flux can
flow. The shift of the critical current distribution towards
higher currents can be described by dependencies of the mean
critical current distribution on the fractal dimension, as it is
shown in the inset of Fig.~\ref{fig11}. The mean critical current
obeys the strong super-linear law specified by Euler gamma
function:
\begin{equation}
\overline{i}=\left( \frac{2+D}{2}\right) ^{\left( 2+D\right) /2}\Gamma
\left( 1-\frac{D}{2}\right)  \label{meancur22}
\end{equation}
Figure \ref{fig11} clearly demonstrates that increasing the
fractal dimension gives a growth of the contribution made by
clusters of greater critical current to the overall distribution,
resulting just in enhancement of the magnetic flux trapping.

\section{ANOMALOUS STATISTICAL PROPERTIES OF THE CRITICAL CURRENT FRACTAL
DISTRIBUTION}

In the general case of gamma-distribution of the normal-phase cluster areas
of Eq.~(\ref{cuma14}) the probability density for the cluster areas has the
form
\begin{equation}
w\left( A\right) =\frac{A^{g}\exp \left( -A/A_{0}\right) }{\Gamma \left(
g+1\right) A_{0}^{g+1}}  \label{probdens23}
\end{equation}
For further consideration it is convenient to introduce the dimensionless
area of the cluster \`{a}, for which the distribution function of Eq.~(\ref
{probdens23}) can be rewritten as:
\begin{equation}
w\left( a\right) =\frac{\left( g+1\right) ^{g+1}}{\Gamma \left( g+1\right) }%
a^{g}\exp \left( -\left( g+1\right) a\right)  \label{prod24}
\end{equation}

The mean dimensionless area of the cluster is equal to unity,
whereas the variance is determined by one parameter only: $\sigma
_{a}^{2}=1/\left( g+1\right) $. The critical current distribution
has the exponential-hyperbolic form of Eq.~(\ref {densi16}), for
which the mean critical current, instead of Eq.~(\ref
{meancur22}), is
\begin{equation}
\overline{i}=G^{\frac{D}{2}}\frac{\Gamma \left( g+1-D/2\right) }{\Gamma
\left( g+1\right) }  \label{mc25}
\end{equation}
The critical current distribution has the skew bell-shaped form with
inherent broad ``tail`` extended over the region of high currents (see Fig.~%
\ref{fig11}). The mean critical current of Eq.~(\ref{mc25}), just
as mean current of Eq.~(\ref {meancur22}) in the special case of
exponential distribution of the normal-phase clusters areas, is
divergent in the range of fractal dimensions $D\geqslant 2(g+1)$.

The gamma distribution of the critical currents of
Eq.~(\ref{densi16} spreads out with some shift to the right as
$g$-parameter decreases. This broadening can be characterized by
the standard deviation of critical currents
\begin{equation}
\sigma _{i}=G^{\frac{D}{2}}\sqrt{\frac{\Gamma \left( g+1-D\right) }{\Gamma
\left( g+1\right) }-\left( \frac{\Gamma \left( g+1-D/2\right) }{\Gamma
\left( g+1\right) }\right) ^{2}}  \label{sd26}
\end{equation}
The standard deviation grows nonlinearly with increase in the
fractal dimension. The peculiarity of the distribution of
Eq.~(\ref{densi16}) is that its variance becomes infinite in the
range of fractal dimensions $D\geqslant g+1$. At the same time,
the mode of the distribution $modef(i)=(G/\theta
)^{D/2}$ remains finite for all possible values of the fractal dimension $%
1\leqslant D\leqslant 2$. The distributions with divergent
variance are known in probability theory - the classic example of
that kind is Cauchy distribution. \cite{hudson} However, such an
anomalous feature of exponential-hyperbolic distribution of
Eq.~(\ref{densi16}) is of special interest, inasmuch as the
current-carrying capability of a superconductor would be expected
to increase in the region of giant variance. Then the statistical
distribution of critical currents has a very elongated ``tail``
containing the contributions from the clusters of the highest
depinning currents.

The reason for divergence of statistical moments of the
exponential-hyperbolic critical current distribution of
Eq.~(\ref{densi16}) consists in the behavior of the cluster area
distribution of Eq.~(\ref {prod24}).The form of this
function takes essentially different shapes depending on the sign of $g$%
-parameter - from the skew unimodal curve at $g>0$ to the
monotonic curve with hyperbolic singularity at zero point at
$g<0$. In the borderline case of $g=0$, which separates these
different kinds of the functions, the
distribution obeys an exponential law. It is just for negative values of $g$%
-parameter that the mean critical current diverges. The
contribution from the clusters of small area to the overall
distribution grows at $g<0$. Since the clusters of small size have
the least number of weak links over a perimeter, they can best
trap the magnetic flux. Therefore, an increase of the part of
small clusters in the area distribution of Eq.~(\ref{prod24})
leads to a growth of the contribution with high depinning currents
made by these clusters in the critical current distribution of
Eq.~(\ref{densi16}). Just as a result of this feature the mean
critical current diverges at $g<0$. Nevertheless, the total area
between the curve of the probability density and the abscissa
remains finite by virtue of the normalization requirement.

Obviously, the proper critical current cannot be infinitely high
as well as the clusters of infinitesimal area do not really exist.
There is the minimum value of the normal-phase cluster area
$A_{m},$ which is limited by the processes of the film growth. So
in YBCO based composites prepared by magnetron sputtering,
\cite{pla2000} the sample value of minimum area of the
normal-phase cluster has been equal to $A_{m}=2070~$nm$^{2}$ at
mean cluster area $\overline{A}=76500~$nm$^{2}$, that corresponds
to the lower bound of
the dimensionless area of the cluster $a_{m}\equiv A_{m}/\overline{A}=0.027$%
. In view of this limitation, we will describe the distribution of the
normal-phase cluster areas by the truncated version of the probability
density of Eq.~(\ref{prod24}):
\begin{equation}
w(a\left| a\geqslant a_{m}\right) =\frac{h\left( a-a_{m}\right) }{1-W\left(
a_{m}\right) }w\left( a\right)  \label{trun27}
\end{equation}
where $h\left( x\right) \equiv \left\{
\begin{array}{cc}
1 & \ \ for\ \ x\geqslant 0 \\
0 & \ \ for\ \ x<0
\end{array}
\right. $ \ is the Heaviside step function, and
\begin{equation}
W\left( a_{m}\right) \equiv \int\limits_{0}^{a_{m}}w\left( a\right) da=\frac{%
\gamma \left( g+1,\left( g+1\right) a_{m}\right) }{\Gamma \left( g+1\right) }
\label{truna28}
\end{equation}
is the truncation degree, which is equal to the probability $\Pr \left\{
\forall a<a_{m}\right\} $ to find the cluster of area smaller than the least
possible value $a_{m}$ in the initial population. In the expression of Eq.~(%
\ref{truna28}) $\gamma \left( \nu ,z\right) $ is the complementary gamma
function.

The expression of Eq.~(\ref{trun27}) gives the conditional distribution of
probability, for which all the events of finding a cluster of area less than
$a_{m}$ are excluded. Thus the truncation provides a natural way to fulfil
our initial assumption that the cluster size has to be greater than the
coherence and penetration lengths. New distribution of cluster areas gives
rise to the truncated distribution of the critical currents:
\begin{equation}
f(i\left| i\leqslant i_{m}\right) =\frac{h\left( i_{m}-i\right) }{1-W\left(
a_{m}\right) }f\left( i\right)  \label{truncur29}
\end{equation}
where $i_{m}\equiv \left( G/\left( g+1\right) a_{m}\right) ^{D/2}$ is the
upper bound of the depinning current, which corresponds to the cluster of
the least possible area $a_{m}$.

Then, instead of Eqs.~(\ref{sd26}) and (\ref{mc25}), the standard deviation
and mean critical current are
\begin{eqnarray}
\sigma _{i}^{\ast } &=&G^{\frac{D}{2}}\sqrt{\frac{\Gamma \left(
g+1-D,\left( g+1\right) a_{m}\right) }{\Gamma \left( g+1,\left(
g+1\right) a_{m}\right) }}
\nonumber \\
&&\overline{-\left( \frac{\Gamma \left( g+1-D/2,\left( g+1\right)
a_{m}\right) }{\Gamma \left( g+1,\left( g+1\right) a_{m}\right) }\right) ^{2}%
}  \label{trunsd30}
\end{eqnarray}

\begin{equation}
\overline{i}^{\ast }=G^{\frac{D}{2}}\frac{\Gamma \left( g+1-D/2,\left(
g+1\right) a_{m}\right) }{\Gamma \left( g+1,\left( g+1\right) a_{m}\right) }
\label{trunmc31}
\end{equation}

For the truncated distribution of Eq.~(\ref{truncur29}) the
possible values of depinning currents are bounded from above by
the quantity $i_{m}$, therefore the mean critical current as well
as the variance do not diverge any more. Both of these
characteristics are finite for any fractal dimensions, including
the case of maximum fractality ($D=2$). The truncation degree is
related to the least possible area of the cluster by the equation
(\ref{truna28}).

The degree of truncation gives the probability measure of the
number of normal-phase clusters that have the area smaller than
the lower bound $a_{m}$ in the initial distribution. If the
magnitude of $W(a_{m})$ is sufficiently small (no more than
several percent), then the procedure of truncation scarcely
affects the initial shape of the distribution and still enables
the contribution from the clusters of zero area (therefore, of
infinitely high depinning current) to be eliminated. It is
interesting to note that the very similar situation occurs in
analyzing the statistics of the areas of fractal islands in the
ocean. \cite{mandelbrot3} The island areas obey the Pareto
distribution, which also has the hyperbolic singularity at zero
point that causes certain of its moments to diverge. For
exponential distribution of the cluster areas, which is valid in
the above-mentioned case of YBCO films, \cite{pla2000} the
probability to find the cluster of area smaller than am in the
sampling is equal to $\Pr \left\{ \forall a<a_{m}\right\} =2.7\%$
only. In principle, the truncation procedure could be made here,
too, but there is no need for that, because the contribution of
infinitesimally small clusters is finite at $g=0$ (and equal to
zero at $g>0$).

As the fractal dimension increases, the critical current
distribution broadens out (the variance grows), moving towards
higher currents (the mean critical current grows, too). This trend
is further enhanced with decreasing $g$-parameter. The most
current-carrying capability of a superconductor should be achieved
when the clusters of small size, which have the highest currents
of depinning, contribute maximally to the overall distribution of
the critical currents. Such a situation takes place just in the
region of giant variance of critical currents. So far, the least
magnitude of $g$-parameter (equal to zero) has been realized in
YBCO composites containing normal-phase clusters of fractal
dimension $D=1.44$. \cite{pla2000} The critical currents of
superconducting films with such clusters are higher than usual.
\cite{tpl2000,tpl99,pss99} It would thus be expected to further
improve the current-carrying capability in superconductors
containing normal-phase clusters, which will be
characterized by area distribution with negative magnitudes of $g$%
-parameter, especially at the most values of fractal dimensions.

\section{CURRENT-VOLTAGE CHARACTERISTICS AND VORTEX DYNAMICS NEAR THE
RESISTIVE TRANSITION}

By virtue of the fact that any motion of the magnetic flux causes the energy
dissipation in superconductors, the question of how such a process could be
prevented, or only suppressed, is of prime practical importance. The study
of resistive state peculiarities leads to the conclusion that the cluster
fractality exerts influence on the electric field induced by the flux
motion. The found distribution of the critical currents allows us to find
the electric field arising from the magnetic flux motion after the vortices
have been broken away from the pinning centers. Inasmuch as each
normal-phase cluster contributes to the total critical current distribution,
the voltage $V$ across a superconductor can be represented as the response
to the sum of effects made by the contribution from each cluster. Such a
response can be expressed as a convolution integral
\begin{equation}
\frac{v}{r_{f}}=\int\limits_{0}^{i}\left( i-i^{\prime }\right) f\left(
i^{\prime }\right) di^{\prime }  \label{conv32}
\end{equation}
where $r_{f}$ is the dimensionless flux flow resistance. The
similar approach is used universally in all the cases where the
distribution of the depinning currents takes place.
\cite{brown,warnes,worden} The subsequent consideration will be
essentially concentrated on the consequences of the fractal nature
of the normal-phase clusters specified by the distribution of
Eqs.~(\ref{cumi15}), (\ref{exhyp18}), so all the problems related
to possible dependence of the flux flow resistance $r_{f}$ on a
transport current will not be taken up here. The voltage across a
superconductor $V$ and flux flow resistance $R_{f}$ are related to
the
corresponding dimensionless quantities $v$ and $r_{f}$ by the relationship: $%
V/R_{f}=I_{c}(v/r_{f})$.

After substitution of the critical current distribution of Eq.~(\ref{densi16}%
) into the expression of Eq.~(\ref{conv32}), upon integration, we get the
final expression for the current-voltage ({\it V-I}) characteristics in the
general case of the gamma-distribution of the cluster areas:

\begin{eqnarray}
\frac{v}{r_{f}} &=&\frac{1}{\Gamma \left( g+1\right)
}\Biggl(i\Gamma \left(
g+1,Gi^{-2/D}\right)   \nonumber \\
&&-G^{D/2}\Gamma \left( g+1-\frac{D}{2},Gi^{-2/D}\right) \Biggr)
\label{gamma33}
\end{eqnarray}

In the special case of exponential cluster area distribution (g=0) the
general formula of Eq.~(\ref{gamma33}) can be simplified:
\begin{equation}
\frac{v}{r_{f}}=i\exp \left( -Ci^{-2/D}\right) -C^{D/2}\Gamma \left( 1-\frac{%
D}{2},Ci^{-2/D}\right)   \label{ci34}
\end{equation}
where $C\equiv \left( \left( 2+D\right) /2\right) ^{2/D+1}$.

In extreme cases of Euclidean clusters and clusters of the most fractality
the expression of Eq.~(\ref{ci34}) can be further transformed:

(i) Clusters of Euclidean boundary ($D=1$ at $g=0$):
\begin{equation}
\frac{v}{r_{f}}=i\exp \left( -\frac{3.375}{i^{2}}\right) -\sqrt{3.375\pi }%
\text{\thinspace erfc}\left( \frac{\sqrt{3.375}}{i}\right)
\label{eucl35}
\end{equation}

where erfc$(z)$ is the complementary error function.

(ii) Clusters of boundary with the maximum fractality ($D=2$ at $g=0$):
\begin{equation}
\frac{v}{r_{f}}=i\exp \left( -\frac{4}{i}\right) +4%
\mathop{\rm Ei}%
\left( -\frac{4}{i}\right)   \label{fract36}
\end{equation}

where $%
\mathop{\rm Ei}%
\left( z\right) $ is the exponential integral function.

\begin{figure}
\includegraphics{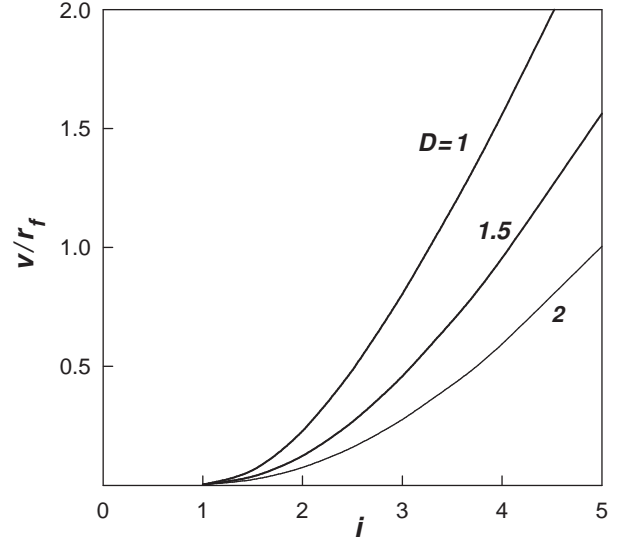}
\caption{\label{fig16} Current-voltage characteristics of
superconductors with fractal clusters of a normal phase.}
\end{figure}

The {\it V-I} characteristics calculated using the formulas of
Eqs.~(\ref {ci34}) - (\ref{fract36}) are presented in
Fig.~\ref{fig16}. Two lines drawn for extreme case of $D=1$ and
$D=2$ bound the region the {\it V-I} characteristics can fall
within for any possible values of fractal dimension. Figure
\ref{fig16} shows that the fractality reduces appreciably an
electric field arising from the magnetic flux motion. This effect
is especially strong in this range of the currents ($1<i<3$),
where the pinning enhancement also has a maximum (see also
Fig.~\ref{fig10}). Both these effects have the same nature,
inasmuch as their reason lies in the peculiarities of the critical
current distribution. The influence of the fractal dimension of
the cluster boundary on the critical current distribution is
demonstrated in Fig.~\ref{fig11}. An increase of
fractality causes a significant broadening of the tail of the distribution $%
f=f(i)$. Therefore, more of small clusters, which have higher critical
currents of depinning, are being involved in the process. Hence the smaller
number of vortices can move, creating the smaller electric field. In turn,
the smaller the electric field is, the smaller is the energy dissipated when
the transport current passes through the sample. Therefore, the decrease in
heat-evolution, which could cause transition of a superconductor into a
normal state, means that the current-carrying capability of a superconductor
containing such fractal clusters is enhanced.

Though the fractality of the clusters reduces the voltage arising
from the vortex motion in the range of the currents $i>1$, the
situation is quite different in the neighborhood of the resistive
transition below the critical current. When $i<1$, the higher the
fractal dimension of the normal-phase cluster is, the larger is
the voltage across a sample and the more stretched is the region
of initial dissipation in {\it V-I} characteristic. The critical
current $i_{c}$ is preceded by some initial region of the finite
voltage drop starting with $i_{on}$, so the resistive transition
is not absolutely abrupt. The existence of this initial section on
the {\it V-I} characteristic arises from the peculiarities of the
fractal distribution in the range of small currents. It has been
just a similar initial region of fractal dissipation has been
observed in high-resolution measurements of dynamical resistance
of HTS-normal metal composites. \cite {prester}

\begin{figure}
\includegraphics{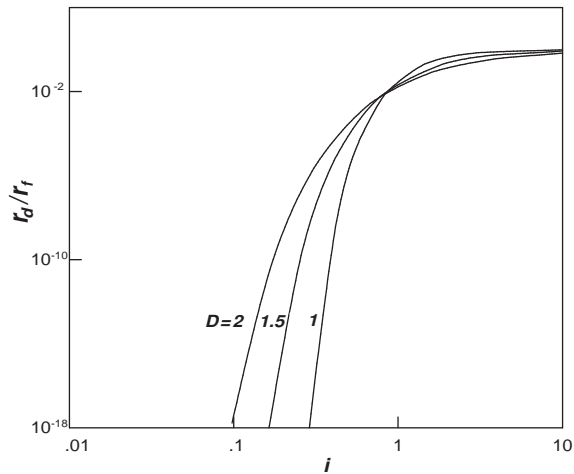}
\caption{\label{fig18} Dependence of the {\it dc} resistance on
the transport current for superconductor with fractal normal-phase
clusters. The dashed horizontal line $r=r_{f}$ at the upper right
corresponds to the viscous flow of a magnetic flux.}
\end{figure}

The significant difference in {\it V-I} characteristic behavior
below and above the resistive transition is related to dependence
of free vortex density on the fractal dimension for various
transport currents. The resistive characteristics provide
additional information about the nature of a vortex state in
superconductor with fractal clusters of a normal phase. Since the
{\it V-I} characteristics of Eqs.~(\ref{gamma33}) - (\ref{ci34})
are non-linear, {\it dc} (static) resistance is not constant and
depends on the transport current. The more convenient parameter is
the differential resistance $r_{d}=dv/di$, a small-signal {\it ac}
parameter that gives the slope of the {\it V-I} characteristic.
The corresponding dimensional quantity $R_{d}$ can be found using
the formulas $R_{d}=r_{d}R_{f}/r_{f}$. Figure \ref{fig18} shows
the graphs of the differential resistance as a function of
transport current for superconductor with fractal normal-phase
clusters. The curves drawn for the Euclidean clusters ($D=1$) and
for the clusters of the most fractal boundaries ($D=2$) bound a
region containing all the resistive characteristics for an
arbitrary fractal dimension. As an example, the dashed curve shows
the case of the fractal dimension $D=1.5$. The dependencies of
resistance on the current shown in Fig.~\ref{fig18} are typical of
the vortex glass, inasmuch as the curves plotted on a double
logarithmic scale are convex and the resistance tends to zero as
the transport current decreases, $r_{d}(i\rightarrow 0)\rightarrow
0$, which is related to the flux creep suppression.
\cite{blatter,brown} A vortex glass represents an ordered system
of vortices without any long-range ordering. At the same time, the
vortex configuration is stable in time and can be characterized by
the order parameter of the glassy state. \cite{fisher,fisher2} In
the {\it H-T} phase diagram, mixed state of the vortex glass type
exists in the region below the irreversibility line. The dashed
horizontal line at the upper right of Fig.~\ref{fig18} corresponds
to a viscous flux flow regime ($r_{d}=r_{f}=$const), which can
only be approached asymptotically.

The resistance of superconductor is determined by the density $n$ of free
vortices broken away from pinning centers of Eq.~(\ref{vort20}). In the
special case of exponential cluster area distribution ($g=0$) the free
vortex density is equal to
\begin{equation}
n=\frac{B}{\Phi _{0}}\exp \left( -C\text{\thinspace
}i^{-2/D}\right) \label{freevort37}
\end{equation}
The more vortices are free to move, the stronger the induced electric field,
and therefore, the higher is the voltage across a sample at the same
transport current. If we differentiate the {\it V-I} characteristic of Eq.~(%
\ref{ci34}) and compare with formula of Eq.~(\ref{freevort37}), we will see
that the differential resistance is proportional to the density of free
vortices: $r_{d}=(r_{f}\Phi _{0}/B)n$. Resistance of a superconductor in the
resistive state is determined by the motion just of these vortices.

\begin{figure}
\includegraphics{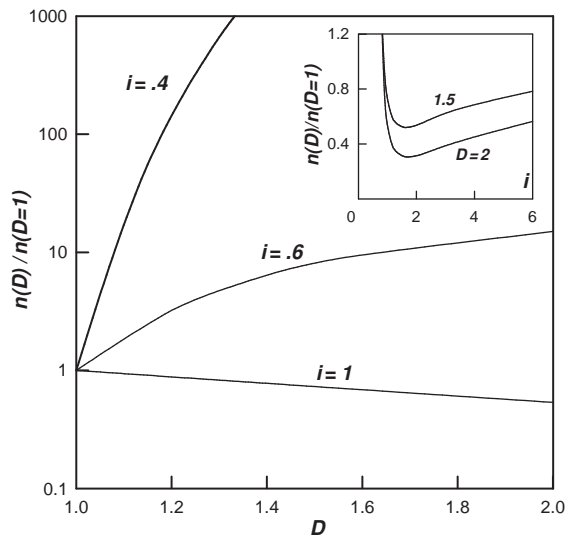}
\caption{\label{fig19} Dependence of the free vortex density on
the fractal dimension of the normal-phase clusters for different
values of a transport current. The inset shows the free vortex
density versus current for two values of a fractal dimensions.}
\end{figure}

Figure \ref{fig19} demonstrates dependence of the relative density
of free vortices $n(D)/n(D=1)$ (relatively to the value for
clusters with Euclidean boundary) on the fractal dimension for
different magnitudes of transport currents. The vortices are
broken away from pinning centers mostly when $i>1$, that is to
say, above the resistive transition. Here the free vortex density
decreases with increasing the fractal dimension. Such a behavior
can be explained by the peculiarity of the critical current
distribution of Eq.~(\ref{densi16}). The curve of this
distribution broadens out, moving towards greater magnitudes of
current as the fractal dimension increases. It means that there
are more clusters of high depinning current in superconductor. The
smaller part of the vortices is free to move, the smaller the
induced electric field. The fact that the fractality of cluster
boundary enhances pinning is also demonstrated in
Fig.~\ref{fig18}, where the resistance decreases with increasing
fractal dimension above the resistive transition. The relative
change in free vortex density depends on the transport current
(see inset in Fig.~\ref{fig19}) and, in the
limiting case of the most fractal boundary $D=2$ reaches a minimum for $%
i=1.6875$ (this curve should goes below others in
Fig.~\ref{fig19}). That corresponds to the maximum pinning gain
and the minimum level of dissipation.

In the range of transport currents below the resistive transition
($i<1$) the resistance as well as the free vortex density increase
for the clusters of greater fractal dimension (see
Figs.~\ref{fig18} and \ref{fig19}). Such a behavior is related to
the fact that the critical current distribution of
Eq.~(\ref{densi16}) broadens out covering both high and small
currents as the fractal dimension increases. For this reason, the
breaking of the vortices away under the action of transport
current begins earlier for the clusters of greater fractal
dimension. In spite of sharp increase in relative density of free
vortices (Fig.~\ref{fig19}), the absolute value of vortex density
in the range of currents involved is very small (much smaller than
above the resistive transition). So the vortex motion does not
lead to the destruction of the superconducting state yet, and the
resistance remains very low. The low density of vortices at small
currents is related to the peculiarity of exponential-hyperbolic
distribution of Eq.~(\ref{densi16}). This function is so ``flat``
in the vicinity of the coordinate origin that all its derivatives
are equal to zero at the point of $i=0$: $d^{k}f(0)/di^{k}=0$ for
any value of $k$. This mathematical feature has a clear physical
meaning: so small a transport current does not significantly
affect the trapped magnetic flux because there are scarcely any
pinning centers of such small critical currents in the overall
statistical distribution, so that nearly all the vortices are
still pinned. This interval corresponds to the so-called initial
fractal dissipation regime, which was observed in BPSCCO samples
with silver inclusions as well as in polycrystalline YBCO and
GdBCO samples. \cite {prester}

\section{Conclusion}

So, in the present work the fractal nature of the normal-phase clusters is
revealed, and relation between the fractal properties of the clusters and
dynamics of the trapped magnetic flux is established. The fractal
distribution of the critical current is obtained. It is found that the
fractality of cluster boundary strengthens the flux pinning and thereby
hinders the destruction of superconductivity by the transport current. {\it %
V-I} characteristics of fractal superconducting structures in a resistive
state are obtained. It is revealed that the fractality of the boundaries of
the normal-phase clusters reduces the electric field arising from magnetic
flux motion, and thereby raises the critical current of a superconductor.
The fractal properties of the normal-phase clusters significantly affect the
resistive transition. This phenomenon is related to the features of the
fractal critical current distribution. The resistance characteristics
correspond to a mixed state of the vortex glass type. An important result is
that the fractality of the normal-phase clusters enhances pinning, thus
decreasing the resistance above the resistive transition and increasing the
current-carrying capability of a superconductor.

\begin{acknowledgments}
This work is supported by the Saint Petersburg Scientific Center
of the Russian Academy of Sciences.
\end{acknowledgments}

\bibliography{FractalWE}

\end{document}